\documentclass[a4paper,fleqn,usenatbib]{mnras}

\usepackage[T1]{fontenc}
\usepackage{ae,aecompl}

\usepackage{graphicx}	
\usepackage{amsmath}	
\usepackage{amssymb}	
\usepackage{subfigure}
\usepackage{color}
\begin{document}
\title[FRB 200428 and the shock synchrotron maser]{The confrontation of the shock-powered synchrotron maser model with the Galactic FRB 200428}
\author[Yu et al.]{Yun-Wei~Yu$^{1,2}$, Yuan-Chuan Zou$^{3}$, Zi-Gao Dai$^{4}$, Wen-Fei Yu$^{5}$
\\$^1$Institute of Astrophysics, Central China Normal
University, Wuhan 430079, China, {yuyw@mail.ccnu.edu.cn}
\\$^2$Key Laboratory of Quark and Lepton Physics (Central
China Normal University), Ministry of Education, Wuhan 430079,
China
\\$^3$School of Physics, Huazhong University of Science and Technology, Wuhan 430074, China
\\$^4$School of Astronomy and Space Science, Nanjing University, Nanjing 210093, China
\\$^5$Key Laboratory for Research in Galaxies and Cosmology, Shanghai Astronomical Observatory, Chinese Academy of Sciences,
\\80 Nandan Road, Shanghai 200030, China}

\maketitle
\begin{abstract}
The association of FRB 200428 with an X-ray burst (XRB) from the Galactic magnetar SGR 1935+2154 offers important implications for the physical processes responsible for the fast radio burst (FRB) phenomena. By assuming that the XRB emission is produced in the magnetosphere, we investigate the possibility that the FRB emission is produced by shock-powered synchrotron maser (SM), which is phenomenologically described with a plenty of free parameters. The observational constraints on the model parameters indicate that the model can in principle be consistent with the FRB 200428 observations, if the ejecta lunched by magnetar activities can have appropriate ingredients and structures and the shock processes occur on the line of sight. To be specific, a complete burst ejecta should consist of an ultra-relativistic and extremely highly collimated $e^{\pm}$ component and a sub-relativistic and wide-spreading baryonic component. The internal shocks producing the FRB emission arise from a collision between the $e^{\pm}$ ejecta and the remnant of a previous baryonic ejecta at the same direction. The parameter constraints are still dependent on the uncertain spectrum and efficiency of the SM emission. While the spectrum is tentatively described by a spectral index of $-2$, we estimate the emission efficiency to be around $10^{-4}$ by requiring that the synchrotron emission of the shocked material cannot be much brighter than the magnetosphere XRB emission.
\end{abstract}

\section{Introduction}
Fast radio bursts (FRBs) are mysterious radio transients, which were usually found at a typical observational frequency around $\sim1$ GHz and have
fluences of a few to a few tens of Jy ms within a time interval of several milliseconds \citep{Lorimer2007,Keane2012,Keane2016,Thornton2013,Burke-Spolaor2014,Spitler2014,
Ravi2015,Masui2015,Champion2016,Caleb2017,Petroff2017,Bannister2017}.  The
anomalously high dispersion measures
(DMs) of FRBs at high Galactic latitudes usually indicate that these phenomena occurred at cosmological distances. According to the millisecond durations and high energy releases of FRBs, their origins are very probably related to the violent activities and even catastrophic collapses/coalescences of compact objects or binaries, in particular, of highly magnetized neutron stars \citep[i.e., magnetars;][]{Popov2010,Kulkarni2014,Katz2016b,Connor2016,Cordes2016,Lyutikov2017}. The magnetar activity model is also strongly supported by the discovery of repeating FRBs (e.g., FRB 121102) and a persistent radio counterpart \citep{Kashiyama2017,Metzger2017,Cao2017b,Dai2017,Michilli2018}.

Very recently, on 28 April 2020, both the CHIME/FRB and STARE2 instruments detected surprisingly an FRB (FRB 200428) from the direction of a Galactic magnetar: SGR 1935+2154 \citep{CHIME2020,Bochenek2020}, while the magnetar had just entered an active state of X-ray bursts (XRBs) since half a month ago \citep{Veres2020,Ridnaia2020a,Barthelmy2020}. On the one hand, the DM of FRB 200428 was measured to be $\sim332.7 \rm pc~cm^{-3}$. On the other hand, SGR 1935+2154 is potentially associated with a supernova remnant G57.2+0.8, which locates in the Outer arm of the Galaxy and has a distance in the range of $6.6-12.5$ kpc \citep{Kothes2018,Zhou2020,Zhong2020}. The consistency between the DM of FRB 200428 and the distance of G57.2+0.8 strengths the magnetar origin of this FRB event.

Specifically, FRB 200428 has a double-peak structure with two components of a width of $\Delta t_{\rm frb}\sim0.5$ ms and separated by $\sim28.91$ ms. The band-averaged fluences of the radio emission were estimated to be 0.7 MJy ms and 1.5 MJy ms in the 400-800 MHz and the 1280-1530 MHz bands, respectively \citep{CHIME2020,Bochenek2020}. Meanwhile, among the detected XRBs, the one occurred at 2020-04-28 14:34:24 UTC was found to be temporally associated with the radio burst, after a correction of the dispersion between the X-ray and radio \citep{CHIME2020,Li2020}. This temporal association robustly suggests the intrinsic connection between the XRB and the FRB. The light curve of this XRB had been well recorded by the SuperAGILE
and Anti-Coincidence detectors \citep{Tavani2020}, the Insight-HXMT \citep{Li2020}, the Konus-Wind \citep{Ridnaia2020b}, and the Integral \citep{Mereghetti2020}, which show that the XRB reached a peak about 0.5s after the trigger and had a duration about one second.

It can almost be certain that FRB 200428 originated from the activity of SGR 1935+2154. Then, it is necessary and interesting to investigate how can the magnetar activity lead to FRB emission. Some theoretical analyses/models have been suggested for understanding this unique FRB event \citep[e.g.,][]{Margalit2020,Lu2020,Dai2020}. As usual, two coherent radiation mechanisms are involved including the curvature radiation and the synchrotron maser (SM) radiation, which have been previously investigated in many literature for cosmological FRBs \citep{Kumar2017,Lu2018,Kumar2020,Yang2018,Yang2020,Wang2019,Lyubarsky2014,Lyutikov2017,Beloborodov2017,Waxman2017,Ghisellini2017,Long2018,Plotnikov2019,Metzger2019}. The focus of this paper is on the SM mechanism powered by a relativistic shock, which is in a debate recently \citep[e.g.][]{Margalit2020,Lu2020}. In Section 2, we briefly describe the observational features of FRB 200428 and its associated XRB. According to these, in Section 3 we carefully derive the observational constraints on the shock-powered SM model, which can provide a basis for judging the availability of the model. Finally, a conclusion and discussions are given in Section 4.

\section{General properties of FRB 200428 }
First of all, by using a reference distance of $d=10$ kpc \citep{Zhong2020}, we can calculate the isotropically-equivalent energy releases of FRB 200428 in the STARE2 frequency range by
\begin{eqnarray}
\mathcal E_{\rm frb}=4\pi d^2 F_{\nu}\Delta\nu=4.5\times10^{34}\rm erg,\label{EFRB}
\end{eqnarray}
where the radio fluence $F_{\nu}=1.5$ MJy ms given by the STARE2 is used and the corresponding $\Delta\nu= 250$ MHz. By contrast, the energy releases from cosmological FRBs were usually found to be around $\sim10^{38}-10^{41}$ erg (without $k$-correction), which makes this Galactic FRB unique. Therefore, in principle, we cannot rule out that FRB 200428 could be intrinsically different from the cosmological ones.

Meanwhile, the average luminosity of FRB 200428 can be calculated by
\begin{eqnarray}
\mathcal L_{\rm frb}=4\pi d^2 S_{\nu}\Delta\nu=7.5\times10^{37}\rm erg~s^{-1},\label{LFRB}
\end{eqnarray}
with $S_{\nu}=2.5$ MJy \citep{Bochenek2020}. This luminosity can be enhanced more, if a wider energy band is taken into account. In comparison, the spin-down luminosity of the magnetar can be estimated by
\begin{eqnarray}
\mathcal L_{\rm sd}={B_{\rm p}^2 R_{\rm s}^6 \Omega^4\over6 c^3}=1.4\times10^{34}\rm erg~s^{-1},\label{Lsd}
\end{eqnarray}
where $B_{\rm p}=4.0\times 10^{14}$ G, $R_{\rm s}\approx10$ km, $\Omega=2\pi/P$, and $P=3.24$ s are the magnetic field strength, the radius, the spin frequency, and the spin period of the magnetar, respectively \citep{Israel2016}. It is showed that, unless the FRB emission is actually collimated within a solid angle smaller than one thousandth of $4\pi$, the spin-down power is not enough to drive the FRB emission. From this point of view, some related FRB models such as the giant pulse model \citep{Connor2016,Cordes2016,Lyutikov2017} could be disfavored. Alternatively, FRB 200428 is very likely to be powered by the engine responsible for the associated XRB, in view of the huge energy release during the XRB as
\begin{eqnarray}
\mathcal E_{X}=4\pi d^2 F_{\rm X}=8.6\times10^{39}\rm erg.\label{EX}
\end{eqnarray}
Here, the unabsorbed fluence of the XRB in the 1-250 keV band is taken as $(7.17^{+0.41}_{-0.38})\times10^{-7} \rm erg/ cm^2$ \citep{Li2020}.

The brightness temperature of FRB 200428 at $\nu\sim1.4$ GHz can be preliminarily estimated by
\begin{eqnarray}
T_{\rm B,1.4GHz} = { c^2\over 2 k_{\rm B} \nu^2}{S_{\nu}d^2 \over \Sigma} \sim 10^{31}\rm K,\label{TB0}
\end{eqnarray}
where the area of emitting region is estimated by $\Sigma=4\pi (c \Delta t_{\rm frb})^2$ with $\Delta t_{\rm frb}\sim 0.5$ ms. This result is much lower than those of the cosmological FRBs but still requires coherent radiation. Therefore, the radio emission must not be the extension of the XRB emission. Instead, it could be another consequence of the magnetar activity in addition to the XRB emission. As a natural consideration, a relativistic ejecta could be driven by the magnetar activity and be shot into the wind region of the magnetar. Then, in principle, it is possible to drive SM radiation by the ejecta, if the ejecta can collide with a magnetized plasma previously existing in the magnetar surroundings.

\begin{figure}
\centering\resizebox{1.0\hsize}{!}{\includegraphics{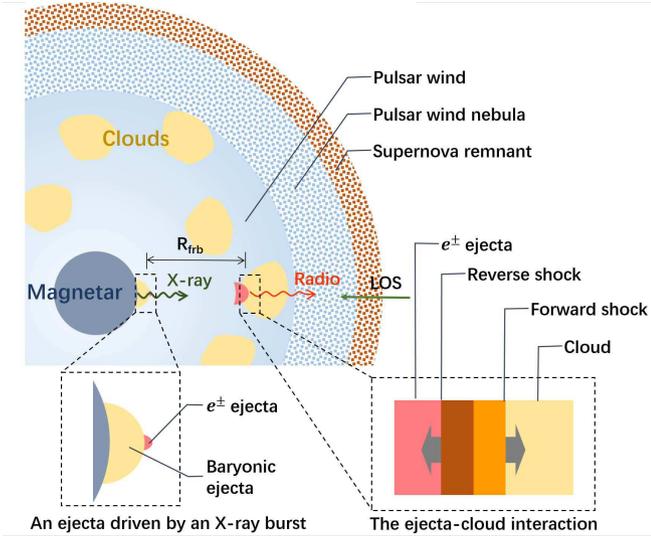}}
\caption{A cartoon illustration of FRB 200428 that arises from a shock-powered SM radiation at a radius of $R_{\rm frb}$. Two shocks are driven by a collision of an ultra-relativistic $\rm e^{\pm}$ ejecta with an ion-electron plasma cloud, where the forward shock is responsible for the radio burst emission. The $\rm e^{\pm}$ ejecta is generated by the XRB temporally associated with FRB 200248, and the pre-existing cloud is just the ejecta remnant (the baryonic component) of a previous XRB. After the reverse shock crosses the $\rm e^{\pm}$ ejecta, the forward shock will be decelerated and then the FRB emission decreases rapidly. So, the intrinsic duration of the FRB emission is primarily determined by the thickness of the $\rm e^{\pm}$ ejecta \citep{Sari1995}. This scenario is different from \citep{Metzger2019} and \citep{Margalit2020}, where the FRB emission is considered to be produced by a forward shock being decelerated in a wind-like medium. }\label{Illustration}
\end{figure}

\section{The model and observational constraints}
The basic assumptions of our model and the involved physical processes are illustrated in Figure \ref{Illustration}. Firstly, a violent activity occurring on/near the magnetar's surface leads to an XRB emission in the magnetar's magnetosphere. Secondly, the activity also drives an ejecta consisting of an ultra-relativistic $e^{\pm}$ component and a sub-relativistic baryonic component. Thirdly, the $e^{\pm}$ ejecta can collide with the remnant of a previous baryonic ejecta at a large radius, which leads to the formation of a pair of internal shocks. The SM emission is powered by the internal forward shock, which can keep constant as long as the reverse shock exists. Fourthly, the SM emission is highly beamed because the shock is highly relativistic and, in particular, the $e^{\pm}$ ejecta is extremely narrow. Consequently, the FRB-XRB association can be detected only when the collision happens on the line of sight (LOS), which significantly suppresses the observational opportunity of such associations. The reasons of these conceptions are analyzed as follows, by setting the parameter values according to the observations.

\subsection{The radiation radius and burst ejecta properties}
It is usually suggested that SM radiation can be powered by a relativistic shock propagating into a moderately magnetized ($\sigma>10^{-3}$) medium \citep{Lyubarsky2014,Beloborodov2017,Metzger2019}. The charges entering the shock gyrates around the ordered magnetic field and then the shock is mediated by the Larmor rotation of the charge. In the magnetar activity model, such a shock can arise from a collision of an ultra-relativistic ejecta with ambient medium. More specifically, the ejecta, which probably consists of electron-positron pairs, can be driven by the engine that triggers the XRB. The ambient medium could either be the magnetar wind nebula \citep{Lyubarsky2014} or a baryon-loaded cloud existing in the magnetar wind \citep{Metzger2019}.

By considering that it may be not intrinsically different from the other XRBs of the magnetar, the XRB associated with FRB 200428 is assumed to occur in the magnetosphere. This is beneficial for understanding the different durations of the FRB and XRB emission. In this case, the casuality between the XRB and the FRB can be connected by the motion of the $\rm e^{\pm}$ ejecta from the magnetosphere to the ambient medium at the radius of $R_{\rm frb}$. This catch-up process leads to an intrinsic time delay for the radio emission relative to the XRB as
\begin{eqnarray}
t_{\rm del}&=&{R_{\rm frb}\over v_{\pm}}\left(1-{v_{\pm}\over c}\cos \theta_{\rm v}\right),\label{tdel}
\end{eqnarray}
where $v_{\pm}$ is the velocity of the $\rm e^{\pm}$ ejecta and $\theta_{\rm v}$ is the viewing angle between the velocity direction and the LOS. The observation of FRB 200428 and its associated XRB shows that this intrinsic time delay cannot be longer than a few millisecond \citep{CHIME2020,Li2020}. In this case, Equation (\ref{tdel}) indicates the motion of the ejecta must be ultra-relativistic of a Lorentz factor $\Gamma_{\pm}$ and the angle $\theta_{\rm v}$ should be smaller than $1/\Gamma_{\pm}$. Therefore, we can get\footnote{In other works \citep[e.g.][]{Margalit2020,Lu2020}, the FRB radiation radius is usually estimated by the formula $R_{\rm frb}= 2\Gamma^2 c \Delta t_{\rm frb}\sim(10^{11}-10^{13})\rm cm$, where $\Gamma\ll\Gamma_{\pm}$ is the Lorentz factor of the shocked region. On the one hand, at such a small radius, the collision between the wide-opening $\rm e^{\pm}$ ejecta and the baryonic cloud can easily happen on the LOS. This indicates the SGR XRBs can often be accompanied by FRB emission, which is however inconsistent with the observations of SGR 1935+2154. Furthermore, according to Equation (\ref{Mc}) and (\ref{Ec}), a small $R_{\rm frb}$ would lead the corresponding mass and energy of the baryonic ejecta to be too small to coincide with the energy scale of the XRB. On the other hand, \cite{Lu2020} suggested that an FRB pulse was produced by a collision between two successive relativistic ejecta. In this case, the widths of the FRB pulses are expected to increase with the time, which is however undiscovered in observation. Therefore, in this paper, we suggest that the $\rm e^{\pm}$ ejecta is a very narrow and even cylindrical jet. So, the variability timescale of the FRB pulses is determined by $\Delta_{\rm frb}\sim l_{\pm}^2/(R_{\rm frb}c)$, where $l_{\pm}$ is the length-scale of the cross section of the $\rm e^{\pm}$ ejecta. In other words, the expression of $R_{\rm frb}=2\Gamma^2 c \Delta t_{\rm frb}$ is inapplicable in our model. This consideration leads the following calculations to be very different from the other works.}
\begin{eqnarray}
R_{\rm frb}&=& 2\Gamma_{\pm}^2 c t_{\rm del}=6.0\times10^{14}\Gamma_{\pm,3.5}^2t_{\rm del,-3}\rm cm.\label{Rfrb}
\end{eqnarray}
For the typical value of $\Gamma_{\pm}$ (see Eq. \ref{parameter4}), the relatively large radius indicates that the ambient medium blocking the $\rm e^{\pm}$ ejecta cannot be the magnetar wind nebula. Instead, it is probably a baryonic cloud, which could be a remnant of the previous burst ejecta.

Therefore, as a complete understanding, the ejecta driven by the mangetar activity should consist of an ultra-relativistic $\rm e^{\pm}$ component and a sub-relativistic baryonic component. The baryonic component could be erupted from the magnetar crust during the violent activity. Violent activities can frequently happen in the whole life of the magnetar and, in particular, intensively in an active state. So, it is expected that many baryonic ejecta can be distributed in the magnetar wind at different radius and different directions, as illustrated in Figure \ref{Illustration}. Here, because of the low luminosity of the magnetar wind, the interaction of the wind with the burst ejecta, including both the leptonic and baryonic components, can be ignored until the ejecta collides with the nebula. Because of their different origins and ingredients, the leptonic and barynic components of the burst ejecta probably have different magnetizations and different durations. It is suggested that the highly magnetized $\rm e^{\pm}$ ejecta can only be produced during the peak of the burst energy release with a millisecond duration, whereas the baryonic ejecta can last for the whole burst period of a few seconds.

\subsection{Parameter constraints}
According to Equation (\ref{tdel}), the XRBs of magnetars can be temporally associated by a radio burst, only if an ion-electron plasma cloud has exist in the moving direction of the new-launched $\rm e^{\pm}$ ejecta. Furthermore, the collision should happen on the LOS. The pre-existing cloud can be erupted from the magnetar earlier than the FRB event by an advance time of $t_{\rm a}\sim {R_{\rm frb}/ v_{\rm ej,B}}$, where $v_{\rm ej,B}$ is the velocity of the cloud (i.e, the baryonic component of the previous ejecta). Please notice that the advance time involved here is not the waiting time between two successive bursts. By considering of the month-long duration of the magnetar activity, we assume the advance time to be on the order of a few days. Then, the relationship between the velocities of the two ejecta components can be written as
\begin{eqnarray}
\Gamma_{\rm \pm,3.5}^{-2}v_{\rm ej,B,9.5}=1.9t_{\rm del,-3}t_{\rm a,5}^{-1}.\label{GammaVej}
\end{eqnarray}

When an $\rm e^{\pm}$ ejecta collides with an ion-electron cloud, a pair of shocks can be driven. As suggested by \cite{Metzger2019}, the FRB emission could be primarily contributed by the forward shock and the reverse shock contribution is nearly negligible, because of the different magnetizations of their upstream material. According to the numerical simulation by \cite{Plotnikov2019}, we know that the peak frequency $\nu_{\rm pk}$ of an SM spectrum powered by the forward shock is about $3\Gamma_{\rm }$ times of the plasma frequency of the cloud material $\nu_{\rm p}=(n_{\rm c}e^2/\pi m_{\rm e})^{1/2}$ \citep{Metzger2019}, where $\Gamma$ is the Lorentz factor of the shock and, approximately, of the radiating shocked material. The value of $\Gamma$ is probably much smaller than that of the unshocked $e^{\pm}$ ejecta $\Gamma_{\pm}$. Then, the particle number density of the cloud can be estimated by
\begin{eqnarray}
n_{\rm c}={\pi m_{\rm e}\over e^2}\left({\nu_{\rm pk}\over 3\Gamma_{\rm }}\right)^2=1.4\times 10^{6}\nu_{\rm pk,8.5}^{2}\Gamma_{\rm 1}^{-2}\rm cm^{-3}.\label{nc}
\end{eqnarray}
According to the jump condition of the forward shock, the comoving energy density of the shocked cloud can be written as $e'_{\rm sh}=4\Gamma^2n_{\rm c}m_{\rm p}c^2$, where the upstream material is considered to be magnetized for a degree of $\sigma\sim1$ and its influence on the jump condition is ignored for simplicity. Meanwhile, the kinetic energy flux carried by the $\rm e^{\pm}$ ejecta can be calculated by\footnote{This energy flux is obtained here by directly considering of the energy conservation, which can also be derived by using the jump condition of the reverse shock as follows. According to the shock jump condition and the equilibrium between the two shocked regions, we can get $e'_{\rm sh}=4\left({\Gamma_{\pm}\over 2\Gamma}\right)^2n'_{\pm}m_{\rm e}c^2$, where $\Gamma_{\pm}/2\Gamma$ represents the Lorentz factor of the reverse shock measured in the comoving frame of the unshocked  $e^{\pm}$ ejecta and $n'_{\pm}$ is the comoving number density of the unshocked $e^{\pm}$ ejecta. Then, the energy flux of the $e^{\pm}$ ejecta can be given by $L_{\pm}=\Gamma_{\pm}^2n'_{\pm}m_{\rm e}c^3\Sigma_{\pm}=\Gamma^2e'_{\rm sh}\Sigma_{\pm}c$.}
\begin{eqnarray}
L_{\pm}=\Gamma^2e'_{\rm sh}\Sigma_{\pm}c=2.5\times10^{39}\nu_{\rm pk,8.5}^2\Gamma_{1}^2\Sigma_{\pm,21}\rm erg~s^{-1},
\end{eqnarray}
where $\Sigma_{\pm}$ represents the cross section of the $\rm e^{\pm}$ ejecta. First of all, this kinetic energy flux is considered to be comparable to the XRB luminosity\footnote{It needs to be emphasized that we do not consider the $\rm e^{\pm}$ ejecta is a result of the XRB emission. Instead, the material ejection and the XRB are considered to be two separated consequences of a same explosive event with equipartition energies. Whereas the XRB emission is likely to be isotropic, the energy flux of the ejecta is highly collimated.}, so the model parameters should satisfy
\begin{eqnarray}
\nu_{\rm pk,8.5}^{2}\Gamma_{1}^{2}\Sigma_{\pm,21} \approx 4.0\mathcal L_{X,40}.\label{constrait1}
\end{eqnarray}
Here, the length-scale of the ejecta cross section can be constrained to be about $l_{\pm}\sim10^{10}-10^{11}$ cm, which is much smaller than $R_{\rm frb}\sim 10^{14}$ cm. This indicates the $\rm e^{\pm}$ ejecta is a very narrow jet and even has a cylindric structure. In this case, the emission angle of the ejecta should be determined by $1/\Gamma$, rather than by the opening angle of ejecta. Meanwhile, the width of the emission pulse is given by $\Delta t_{\rm frb}\sim l_{\pm}^2/(R_{\rm frb}c)$, rather than by  $\Delta t_{\rm frb}\sim R_{\rm frb}/(2\Gamma^2c)$.

Therefore, on the one hand, the brightness temperature of the FRB should be re-estimated by
\begin{eqnarray}
T_{\rm B,1.4GHz} &=& { c^2\over 8\pi k_{\rm B} \nu^2}{\mathcal L_{\rm frb} \over \Delta \nu\Gamma^2\Sigma_{\pm}} \nonumber\\&=& 5.3\times10^{22}\mathcal L_{\rm frb,37}\Gamma_{1}^{-2}\Sigma_{\pm,21}^{-1}\rm K,
\end{eqnarray}
instead of Equation (\ref{TB0}). The radio burst emission can penetrate the cloud, only if the optical depth due to the induced Compton scattering is smaller than $\sim3$, which can be determined as \citep{Lyubarsky2016,Metzger2019}
\begin{eqnarray}
\tau_{\rm ic,1.4GHz}&=&{1\over10}{k_{\rm B}T_{\rm B}\over m_{\rm e}c^2}\sigma_{\rm T}n_{\rm c}c\Delta t_{\rm frb}\nonumber\\
&=&24.5\mathcal L_{\rm frb,37}\Delta t_{\rm frb,-3}\nu_{\rm pk,8.5}^2\Gamma_{1}^{-4}\Sigma_{\pm,21}^{-1}.
\end{eqnarray}
Then the second constraint on the model parameters can be yielded to
\begin{eqnarray}
\nu_{\rm pk,8.5}^{-2}\Gamma_{1}^{4}\Sigma_{\pm,21}\approx 8.2\mathcal L_{\rm frb,37}\Delta t_{\rm frb,-3}.\label{constrait2}
\end{eqnarray}
On the other hand, if we tentatively take an index of $\alpha=-2$ for the $L_{\nu}\propto\nu^{\alpha}$ spectrum above the peak frequency, then the FRB luminosity can be corrected from the isotropically-equivalent observational one to
\begin{eqnarray}
L_{\rm frb,cor}&\approx &\nu_{\rm pk}{\mathcal L_{\rm frb} \over \Delta \nu\Gamma^2}\left({1.4\rm GHz\over \nu_{\rm pk}}\right)^2\nonumber\\
&=& 2.5\times10^{36}\mathcal L_{\rm frb,37}\nu_{\rm pk,8.5}^{-1}\Gamma_{1}^{-2}\rm erg~s^{-1}.
\end{eqnarray}
In comparison with the total energy flux of the $e^{\pm}$ ejecta, the radiation efficiency of the SM can be constrained to
\begin{eqnarray}
\xi\approx{L_{\rm frb,cor}\over L_{\pm}}\approx1.0\times10^{-3}\mathcal L_{\rm frb,37}\nu_{\rm pk,8.5}^{-3}\Gamma_{1}^{-4}\Sigma_{\pm,21}^{-1}.
\end{eqnarray}
If we require this radiation efficiency to be not much higher than $10^{-4}$ for an upstream magnetization of $\sigma\sim1.0$ \citep{Plotnikov2019}, then we can get the third constraint on the model parameters as
\begin{eqnarray}
\nu_{\rm pk,8.5}^3\Gamma_{1}^{4}\Sigma_{\pm,21}\approx 10\mathcal L_{\rm frb,37}\xi_{-4}^{-1}.\label{constrait3}
\end{eqnarray}
Here the primary uncertainty comes from the uncertain spectral index of the SM spectrum.

For a self-consistency between Equations (\ref{constrait1}),  (\ref{constrait2}), and (\ref{constrait3}), we can finally obtain a plausible set of the model parameters as
\begin{eqnarray}
\nu_{\rm pk}\approx3.3\times10^{8}\Delta t_{\rm frb,-3}^{-1/5}\xi_{-4}^{-1/5}\rm Hz,\label{parameter1}
\end{eqnarray}
\begin{eqnarray}
\Gamma\approx15.4\mathcal L_{X,40}^{-1/2}\mathcal L_{\rm frb,37}^{1/2}\Delta t_{\rm frb,-3}^{1/10}\xi_{-4}^{-2/5},\label{parameter2}
\end{eqnarray}
\begin{eqnarray}
\Sigma_{\pm}\approx1.6\times10^{21}\mathcal L_{X,40}^{2}\mathcal L_{\rm frb,37}^{-1}\Delta t_{\rm frb,-3}^{1/5}\xi_{-4}^{6/5}\rm cm^2,\label{parameter3}
\end{eqnarray}
which are directly dependent on the observational luminosities of the FRB and XRB and as well as the presumed SM radiation efficiency.

For the model parameters obtained above, we can derive a particle number flux of the $e^{\pm}$ ejecta as
\begin{eqnarray}
\dot{N}_{\pm}&=&{L_{\pm}\over(1+\sigma_{\pm})\Gamma_{\pm}m_{\rm e}c^2}\nonumber\\
&=&1.3\times10^{32}(1+\sigma_{\pm})^{-1}\mathcal L_{X,40}\Gamma_{\pm,3.5}^{-1}\rm ~s^{-1},\label{number}
\end{eqnarray}
In principle, this can be explained by the Goldreich-Julian flux of the magnetar as $\dot{N}_{\rm GJ}\sim  \mu_{\pm}f_{\rm b}R_{\rm L}^2n_{\rm GJ}c\sim10^{31}\mu_{\pm}f_{\rm b}\rm s^{-1}$, if the magnetization of the flux $\sigma_{\pm}\gg 1$, the pair multiplicity $\mu_{\pm}\gg1$, and the beaming factor of ejecta $f_{\rm b}\ll1$ have appropriate values, where $R_{\rm L}=c/\Omega$ is the light cylinder radius and $n_{\rm GJ}=(\Omega B_{\rm p}/2\pi ec)(R_{\rm L}/R_{\rm s})^{-3}$ is the Goldreich-Julian density. Furthermore, we can also make an estimation for the cloud properties. Firstly, the length-scale of the cloud in the longitude direction can be estimated by $R_{\rm frb}(r_{\rm b}/R_{\rm s})$, where $r_{\rm b}$ is the radius of the region on the magnetar surface where the baryonic material erupted from. Secondly, the length-scale in the latitude direction can be determined by the ratio of the XRB duration ($\Delta t_{\rm X}\sim1$ s) to the spin period ($\sim3$ s): $\sim 2\pi R_{\rm frb}/3$. Finally, the thickness of the cloud is given by $v_{\rm ej,B}\Delta t_{\rm X}$. Following these considerations, the mass of the cloud and its kinetic energy can be calculated by
\begin{eqnarray}
M_{\rm c}&\approx&{4\pi r_{\rm b}\over3R_{\rm s}} R_{\rm frb}^2v_{\rm ej,B}\Delta t_{X}n_{\rm c}m_{\rm p}\nonumber\\
&=&5.0\times10^{20} \mathcal L_{X,40}\mathcal L_{\rm frb,37}^{-1}\Delta t_{X,0}\Delta t_{\rm frb,-3}^{-3/5} t_{\rm del,-3}^2 \nonumber\\
&&\xi_{-4}^{2/5}r_{\rm b,5}\Gamma_{\pm,3.5}^4 v_{\rm ej,B,9.5}~\rm g,\label{Mc}
\end{eqnarray}
and
\begin{eqnarray}
E_{\rm ej,B}&=&{1\over 2}M_{\rm c}v_{\rm ej,B}^2\nonumber\\
&=&2.5\times10^{39} \mathcal L_{X,40}\mathcal L_{\rm frb,37}^{-1}\Delta t_{X,0}\Delta t_{\rm frb,-3}^{-3/5} t_{\rm del,-3}^2 \nonumber\\
&&\xi_{-4}^{2/5}r_{\rm b,5}\Gamma_{\pm,3.5}^4 v_{\rm ej,B,9.5}^3~\rm erg,\label{Ec}
\end{eqnarray}
respectively. By assuming $E_{\rm ej,B}\approx\mathcal E_{X}$ and combining with Equation (\ref{GammaVej}), we can get
\begin{eqnarray}
\Gamma_{\pm}&\approx&2.9\times10^{3} \mathcal L_{\rm frb,37}^{1/10}\Delta t_{\rm frb,-3}^{3/50} t_{\rm del,-3}^{-1/2} \nonumber\\
&&\times\xi_{-4}^{-1/25}r_{\rm b,5}^{-1/10}t_{\rm a,5}^{3/10},\label{parameter4}
\end{eqnarray}
and
\begin{eqnarray}
v_{\rm ej,B}&\approx&5.4\times10^{9} \mathcal L_{\rm frb,37}^{1/5}\Delta t_{\rm frb,-3}^{3/25} \nonumber\\
&&\times\xi_{-4}^{-2/25}r_{\rm b,5}^{-1/5}t_{\rm a,5}^{-2/5}\rm cm~s^{-1},\label{parameter5}
\end{eqnarray}
which are very insensitive to the uncertain model parameters (i.e., $\xi$, $r_{\rm b}$, and $t_{\rm a}$).
As shown, the mass of the baryonic ejecta could be negligible in comparison to the total mass of a magnetar crust ($\sim10^{-5}M_{\odot}$) and thus the production of this baryonic ejecta is possible.

In summary, as listed in Table \ref{table1}, seven crucial model parameters can well be determined by the combination of the five observational quantities and four extra model parameters, where the primary uncertainty comes from the physical description of the SM process. These self-consistent parameter constraints suggest that the shock-powered SM model could in principle work for the observations of FRB 200428, if the ejecta has the appropriate ingredients and structures and the SM radiation has an appropriate efficiency. As a further attempt, we discuss more on the value of the radiation efficiency $\xi$ in Section 3.3.

\begin{table}
\centering\caption{Observational quantities and parameter constraints}\label{table1}
\begin{tabular}{c|c|c|c}
\hline\hline
Symbol&Quantity and parameter& Reference value\\
\hline\hline
&\textit{Observational quantities}& \\
\hline
$t_{\rm del}$&Time delay of the FRB to the XRB&$< 10^{-3}$ s\\
$\mathcal L_{\rm frb}$&FRB luminosity&$\sim 10^{37}\rm erg~s^{-1}$\\
$\Delta t_{\rm frb}$&FRB width&$\sim 10^{-3}$ s\\
$\mathcal L_{\rm X}$&XRB luminosity&$\sim 10^{40}\rm erg~s^{-1}$\\
$\Delta t_{\rm X}$&XRB width&$\sim 1$ s\\
\hline\hline
&\textit{Tentative parameters}& \\
\hline
$t_{\rm a}$&Time interval between two burst& $\sim1$ day\\
& activities on the same direction& \\
$r_{\rm b}$&Radius of the burst region & $\sim1$ km\\
& on the magnetar surface& \\
$\alpha$&Spectral index of the SM emission&$-2$&\\
$\xi$&Efficiency of the SM emission&$\sim10^{-4}$&\\
\hline\hline
&\textit{Parameter constraints}& \\
\hline
$\nu_{\rm pk}$&Peak frequency of the SM emission&Eq. (\ref{parameter1})&\\
$\Gamma$&Lorentz factor of the shocked region&Eq. (\ref{parameter2})&\\
$\Sigma_{\pm}$&Cross section of the $e^{\pm}$ ejecta &Eq. (\ref{parameter3})&\\
$\dot{N}_{\pm}$&Particle number flux of the $e^{\pm}$ ejecta &Eq. (\ref{number})&\\
$M_{\rm c}$&Mass of the baryonic ejcta &Eq. (\ref{Mc})&\\
$\Gamma_{\pm}$&Lorentz factor of the $e^{\pm}$ ejecta& Eq. (\ref{parameter4})&\\
$v_{\rm ej,B}$&Velocity of the baryonic ejcta&Eq. (\ref{parameter5})&\\
\hline\hline
\end{tabular}
\end{table}

\subsection{Synchrotron emission}
Besides the maser radiation, the shocked material can also release its internal energy through synchrotron emission, which then provides an extra counterpart for the FRB emission. As usual, the synchrotron emission can be characterized by the following two characteristic frequencies
\begin{eqnarray}
\nu_{\rm m} &=&  {3 e{B'}_{\rm sh}\over4 \pi m_{\rm e}c} \gamma_{\rm m}^2\Gamma\nonumber\\&=&1.3\times10^{17} \mathcal L_{X,40}^{-3/2}\mathcal L_{\rm frb,37}^{3/2}\Delta t_{\rm frb,-3}^{1/10}\xi_{-4}^{-7/5} \sigma_{0}^{1/2}\epsilon_{\rm e,-1}^{2} \rm Hz,
\end{eqnarray}
\begin{eqnarray}
\nu_{\rm c} &=&  {3 e{B'}_{\rm sh}\over4 \pi m_{\rm e}c} \gamma_{\rm c}^2\Gamma\nonumber\\&=&1.5\times10^{18}\mathcal L_{X,40}^{1/2}\mathcal L_{\rm frb,37}^{-1/2}\Delta t_{\rm frb,-3}^{1/2}\xi_{-4}^{} \sigma_{0}^{-3/2}t_{-3}^{-2}\rm Hz,
\end{eqnarray}
and the chromatic luminosity at the peak frequency $\min(\nu_{\rm m},\nu_{\rm c})$
\begin{eqnarray}
L_{\nu,\max}^{\rm syn} &=&  N_{\rm e,sh}{m_{\rm e}c^2 \sigma_{\rm T}\over3 e}{B'}_{\rm sh} \Gamma\nonumber\\&=&3.9\times10^{20}\mathcal L_{X,40}^{3/2}\mathcal L_{\rm frb,37}^{-1/2}\Delta t_{\rm frb,-3}^{-3/10}\nonumber\\
&&\times\xi_{-4}^{1/5} \sigma_{0}^{1/2}t_{-3}\rm erg~s^{-1}Hz^{-1},
\end{eqnarray}
where $\gamma_{\rm m} = \epsilon_{\rm e} \Gamma{(m_{\rm p}/ m_{\rm e})} {(p - 2)/(p - 1)}$, $\gamma_{\rm c} =  {6 \pi m_{\rm e}c/( \sigma_{\rm T}{B'}_{\rm sh}^2\Gamma t)}$, ${B'}_{\rm sh}=\sqrt{8\pi \sigma{e'}_{\rm sh}}$, and the total number of the shocked electrons is given by $N_{\rm e,sh}=2\Gamma^2 ct\Sigma_{\pm} n_{\rm c}$. The parameters $\epsilon_{\rm e} $ and $p$ are introduced to describe the energy fraction and the distribution index of the shocked electrons. The dynamical time $t$ of the shock is on the order of a millisecond, by considering that the $e^{\pm}$ ejecta has a millisecond duration. For the adopted parameter values, the shocked electrons in the slow cooling state can determine a luminosity of
\begin{eqnarray}
L_{X}^{\rm syn} &=&  {\nu_{\rm c}L_{\nu,\max}^{\rm syn}}\left(\nu_{\rm c}\over\nu_{\rm m}\right)^{-(p-1)/2}\nonumber\\&=&1.2\times10^{38}\mathcal L_{X,40}^{(3-p)}\mathcal L_{\rm frb,37}^{p-2}\Delta t_{\rm frb,-3}^{(2-p)/5}\nonumber\\
&&\times\xi_{-4}^{6(2-p)/5}\epsilon_{\rm e,-1}^{p-1} \sigma_{-1}^{p-2}t_{-3}^{p-2}\rm erg~s^{-1}cm^{-2}\rm erg~s^{-1},
\end{eqnarray}
which seems much smaller than the isotropic luminosity of the original XRB. However, by considering of the relativistic beaming of the shock emission, the observed flux of this synchrotron X-ray emission can actually reach to be as high as
\begin{eqnarray}
F_{X}^{\rm syn} &=&  {\Gamma^2 L_{X}^{\rm syn}\over 4\pi d^2}\nonumber\\&=&2.4\times10^{-6}\mathcal L_{X,40}^{2-p}\mathcal L_{\rm frb,37}^{p-1}\Delta t_{\rm frb,-3}^{(3-p)/5}\nonumber\\
&&\times\xi_{-4}^{2(4-3p)/5}\epsilon_{\rm e,-1}^{p-1} \sigma_{0}^{p-2}t_{-3}^{p-2}\rm erg~s^{-1}cm^{-2},\label{FX2}
\end{eqnarray}
which is even several times higher than the flux of the original XRB. This intense millisecond X-ray pulse overlapping the XRB emission provides an observational signature of the shock-powered SM model, although it could be not very easy to be identified from the X-ray observations because of its short duration. In any case, it is indicated that the value of $\xi$ cannot be much smaller than $10^{-4}$, otherwise this synchrotron X-ray pulse would become too significant to be consistent with the current observation. This requirement further consolidates the observational constraints presented in Section 3.2, in particular, the values of the parameters $\sigma$ and $\nu_{\rm pk}$ that are usually adopted as assumptions in other works.

In principle, as another possibility, it cannot be ruled out that, maybe, the observed XRB emission itself is actually contributed by this shock synchrotron emission \citep{Margalit2020}, rather than being produced in the magnetosphere. In this case, for explaining the different temporal behaviors of the FRB and XRB, it is required that the shock can at least last for a time much longer than the deceleration timescale \citep{Metzger2019}. The material blocking the $e^{\pm}$ ejecta should be wind-like rather than shell-like, but it is actually unclear how can such a wind-like environment is formed via the intermittent burst activities of the magnetar.


\section{Conclusion and discussions}
The discovery of FRB 200428 from the Galactic magnetar SGRB 1935+2154 during its active state robustly indicated that the FRB phenomena can be produced by magnetars and, more specifically, be powered by the burst activities. Then a basic question is how the FRB emission can be generated during an XRB, which attracts wide interesting and leads to a  debate. This paper is devoted to test the applicability of the shock-powered SM model in accounting for this FRB-XRB association event. It is found that the model can match the observations in best if the following crucial conditions can be satisfied:
\begin{itemize}
\item The FRB emission is produced by the internal shocks between an $e^{\pm}$ ejecta and a previous baryonic ejecta, rather than produced by an external shock decelerating in a wind-like medium \citep{Metzger2019,Margalit2020} or by the internal shocks between two relativistic ejecta \citep{Lu2020}. In this case, the FRB emission would not evolve with time as observed.
\item The XRB emission is produced in the magnetosphere of the magnetar, rather than by the shock responsible for the FRB emission \citep{Margalit2020}. Then, it is natural to expect the XRB and the FRB emission can have different durations. The shock synchrotron X-ray emission overlapping the magnetosphere XRB emission can somewhat make the FRB-associated XRB different from the other normal ones. This emission characteristic provides a signature to test the shock-powered SM model.
\item The radiation radius of the FRB is required to be as large as $\sim 10^{14}$ cm. Meanwhile, it is inferred that the $e^{\pm}$ ejecta is a very narrow jet of a cross section of $\sim10^{21}\rm cm^2$, rather than a wide-spreading outflow. This fact leads to the rarity of the FRB-XRB association, just as constrained by the FAST observations \citep{Lin2020}, since the association can be detected only if the collision occurs on the LOS.
\end{itemize}
Because of the extremely high collimation of the $e^{\pm}$ ejecta, the traditional estimate of the radiation radius $\sim2\Gamma^2 c\Delta t_{\rm frb}$ becomes unavailable. Instead, the widths of the FRB pulses are actually determined by the dynamical timescale of the crossing time of the internal reverse shock and the time delay of the high-latitude emission of the narrow jet.

If the collision is observed off-axis, a relatively long time delay would appear between the FRB and the magnetosphere XRB. Furthermore, the FRB emission can be significantly reduced by the relativistic beaming effect. Nevertheless, if the deviating angle of the collision from the LOS is not too large (e.g., $\Gamma^{-1}<\theta_{\rm v}\ll1$), a relatively weak FRB could still be detected. The corresponding time delay can be given by $\Delta t_{\rm frb}\sim R_{\rm frb}\theta_{\rm v}^2/( 2c)$, which makes it not easy to find out the FRB-XRB association. The relatively faint radio pulse detected by FAST on 30 April \citep{Zhang2020} could just be in such a case. Additionally, of course, we cannot rule out that this faint radio pulse could in fact belong to the pulsation radio emission of the magnetar.

In spite of the above self-consistent description of the model, an uncertainty can still come from the most crucial but very unclear physics of the relativistic-shock-powered SM emission, including its spectral shape and radiation efficiency. By taking a tentative spectral index, in this paper we empirically obtain an observational constraint on the peak frequency of the spectrum and the radiation efficiency, which at least indicates the model can be in principle viable. In any case, in the future, basic principles calculations are still definitely necessary for describing the fundamental physics of the shock-powered SM process and testing its applicability in the FRB phenomena.

Additionally, the physical processes described in this Letter make this Galactic FRB somewhat similar to type III solar radio bursts, which can also be associated with some X-ray flares \citep{Pick2008} and can be accounted for by an energy transfer from an electron beam to the ambient medium \citep{Reid2014}. Such a similarity between FRBs and solar radio bursts had also been found by a statistical study for the cosmological repeating FRBs by \citep{Zhang2019}. These somewhat provide circumstantial evidences supporting the shock-powered SM model for FRBs.

\section*{acknowledgements}
The authors thank Bing Zhang for useful discussions. This work is supported by the
National Natural Science Foundation of China (Grant Nos.
11822302 and 11833003) and the Fundamental
Research Funds for the Central Universities (Grant
No. CCNU18ZDPY06).

\section*{data availability}

No new data were generated or analysed in support of this research.

\end{document}